\documentclass[12pt]{article}

\usepackage[a4paper,margin=1in]{geometry}
\usepackage{setspace}
\usepackage{graphicx}
\usepackage{amsmath}
\usepackage{adjustbox}  
\usepackage{tabularx}
\usepackage{booktabs}
\usepackage{ragged2e}
\usepackage{ltablex}
\keepXColumns
\usepackage{svg}
\svgpath{{figures/}}

\usepackage[numbers,sort&compress]{natbib}
\usepackage{hyperref}
\hypersetup{colorlinks=true, linkcolor=blue, citecolor=blue, urlcolor=blue}
\usepackage[affil-it]{authblk}   

\title{Virtual Cells: From Conceptual Frameworks to Biomedical Applications}

\author[1,5]{Saurabh Bhardwaj \thanks{These authors contributed equally to this work.}}
\author[1]{Gaurav Kumar\textsuperscript{*}}
\author[2]{Haochen Yang\textsuperscript{*}} 
\author[3]{Shaurya Bhardwaj}
\author[4]{Qun Wang}
\author[5]{Minjie Shen}
\author[5]{Yizhi Wang}
\author[6]{Cristabelle Madona De Souza\textsuperscript{*}\thanks{Correspondence: \texttt{cdesouza@stanford.edu, saurabh.bhardwaj@thapar.edu}}}

\affil[1]{Department of Electrical \& Instrumentation Engineering, Thapar Institute of Engineering \& Technology, India}
\affil[2]{School of Engineering \& Applied Sciences, Harvard University, USA}
\affil[3]{Department of Computer Science \& Engineering, Thapar Institute of Engineering \& Technology, India}
\affil[4]{Department of Computer Science, San Francisco State University, USA}
\affil[5]{Department of Electrical \& Computer Engineering, Virginia Tech, USA}
\affil[6]{Department of Radiation Oncology, Stanford University, USA}
\date{}
\begin{document}
\maketitle
\vspace{-1.5cm} 

\begin{abstract}
The challenge of translating vast, multimodal biological data into predictive and mechanistic understanding of cellular function is a central theme in modern biology. Virtual cells, or digital cellular twins, have emerged as a critical paradigm to meet this challenge by creating integrative computational models of cellular processes. This review synthesizes the evolution and current state of the virtual cell, from foundational mechanistic frameworks like the Virtual Cell that employ deterministic and stochastic simulations to the recent transformative impact of artificial intelligence and foundation models. We examine the core technological pillars required to build these models, including the integration of various data types, such as single-cell and spatial omics, the spectrum of modeling approaches, and the bioengineering principles that connect simulation to application. We further discuss key applications, frameworks for model benchmarking and validation, and the significant hurdles that remain, including computational scalability, parameter inference, and ethical considerations, which provides a roadmap for the future development of predictive virtual cells that promise to revolutionize biomedical research and clinical practice.\\

\end{abstract}

\vspace{-1.5cm} 

\section{Introduction}

\begin{figure*}[t]
  \centering
  \includegraphics[width=\textwidth]{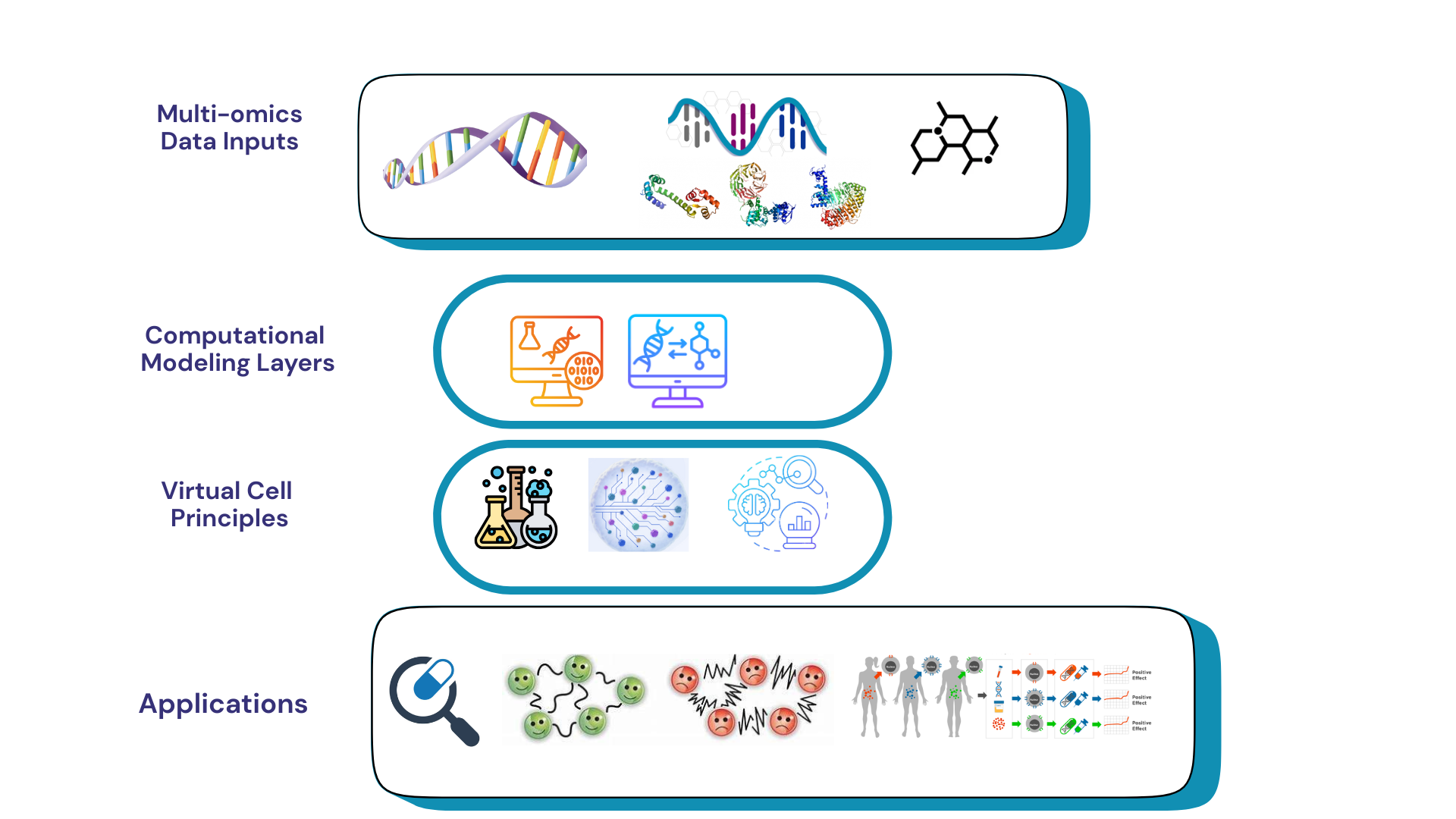}
  \caption{Conceptual framework of Virtual Cells integrating multi-omics data, computational modeling, core principles, and applications for systems biology}
  \label{fig:conceptual-framework-wide}
\end{figure*}

The rapid growth of high-throughput technologies has generated unprecedented volumes of multi-omics data, creating both opportunities and challenges for modern biology. Translating these complex data sets into actionable biomedical knowledge is crucial to advance drug discovery, disease modeling, and precision medicine, as illustrated in Figure~\ref{fig:conceptual-framework-wide}. However, conventional reductionist approaches often fall short in representing the non-linear, dynamic, and multi-scale interactions that define cellular processes. Cells operate as integrated systems of genes, proteins, metabolites, and signaling pathways, where emergent behaviors cannot be fully explained by studying individual components. This gap underscores the need for integrative computational frameworks, such as the Virtual Cell, that can bridge experimental biology with predictive, system-level modeling.

The conceptual foundations of Virtual Cell are linked to the development of synthetic biology, where initial research highlighted the necessity of applying engineering principles to biological design.  Initial contributions laid the groundwork for modularity, standardization, and translational relevance, which remain fundamental concepts influencing contemporary computational frameworks.
The progression of synthetic biology has created the theoretical and practical foundations required for the development of Virtual Cell frameworks. Drew Endy highlighted the lack of engineering concepts as a significant obstacle to advancement in the field \cite{endy2005foundations}. The idea outlines standardization, decoupling, and abstraction as essential methods, offering a framework for modular and reproducible biological design.Wendell highlighted modularity as a crucial characteristic of designed systems, promoting the concept of cells operating as programmable units via the incorporation of gene circuits into cellular decision-making mechanisms \cite{lim2022emerging}. The integration of these perspectives fosters modularity and repeatability as fundamental design principles that align closely with the architecture of Virtual Cell frameworks . \\
Historical investigations provide further background for this trend. The early researchers carried out a comprehensive review of the evolution of synthetic biology, highlighting its shift from informal genetic modification to a field grounded in systematic design principles \cite{cameron2014brief}. Their work highlights the importance of establishing formal standards and frameworks, which provided the foundation for computational methods like Virtual Cell. Fischbach et.al. stated that modified cells signify a novel basis for medicine, enhancing existing methods like pharmaceuticals and technologies \cite{fischbach2013cell}. Their viewpoint, while primarily focused on therapeutic translation, highlights the necessity for predictive and design-oriented modeling. The contributions collectively underscore the alignment of essential concepts such as modularity, standardization, and translational potential in shaping the emerging idea of the Virtual Cell. \\
Virtual Cells represent an emerging frontier in cell biology, integrating multi-omics data and computational modeling to replicate cellular processes in silico. Recent developments in multimodal modeling, benchmarking challenges, and practical biomedical applications have positioned Virtual Cells as crucial tools in predictive biology, personalized medicine, and drug discovery\cite{singhvi1994engineering},\cite{endy2005foundations}.  Nevertheless, the scale and complexity of biological systems necessitate methodologies that surpass reductionist biology.\\
In this context, the idea of the Virtual Cell has become a new way of thinking about things.  Virtual Cells are computational models of biological cells that combine multi-omics data, molecular networks, and systems-level interactions into predictive, dynamic models \cite{kalos2011t}.  Virtual Cells enable the simulation of cellular processes in silico, which promises to hasten biomedical discovery, diminish dependence on expensive wet-lab experimentation, and yield mechanistic insights that are challenging to acquire through conventional experimental approaches \cite{czi2025cz}.  
  
At its foundation, the concept of a Virtual Cell is aimed at representing the fundamental biochemical, genetic, and physiological processes of living systems through a combination of mathematical modeling, network reconstruction, and computational simulation \cite{jinek2012programmable}, \cite{cong2013multiplex}. Unlike static frameworks, these models are inherently dynamic, incorporating temporal changes, spatial organization \cite{fan2025large}, and stochastic variations that collectively define cellular behavior in real biological contexts \cite{rozenblatt2017human}. Different modeling approaches fall under the umbrella of Virtual Cells. For example, metabolic models focus on enzymatic pathways and flux distributions, while regulatory models address mechanisms of gene expression and transcriptional regulation. Similarly, signaling models are employed to capture cascades that mediate how cells perceive and respond to environmental cues, and whole-cell models strive to integrate multiple biological layers into a unified representation that approximates overall cellular functionality \cite{ponten2008human}. \\
The significance of Virtual Cells extends beyond their descriptive capacity. They serve as computational counterparts, or digital twins~\cite{hao2024ai}, of living cells, allowing researchers to predict how biological systems respond to genetic alterations, environmental perturbations, or pharmacological treatments \cite{cong2013multiplex}. Through predictive modeling, Virtual Cells facilitate hypothesis generation, enable in silico experimentation, and support the rational design of interventions prior to laboratory testing. This capability not only accelerates discovery but also substantially reduces the costs and resources traditionally required for experimental validation \cite{kalos2011t}.The inception of Virtual Cells can be attributed to the emergence of mathematical biology in the mid-20$^{\text{th}}$ century, during which differential equations were initially utilized to characterize enzyme kinetics and population dynamics \cite{Cho2022OpenCell},\cite{Thul2017SubcellularMap}.  The emergence of systems biology in the late 1990s signified a pivotal shift, highlighting comprehensive methodologies that incorporated molecular networks instead of discrete routes \cite{czi2025cz}, \cite{Specht2021SCoPE2}  The Virtual Cell project and other early platforms created the first computing systems for simulating cellular processes, showing that mechanistic cell models could work \cite{singhvi1994engineering}.

\begin{figure*}[t]
  \centering
  \includegraphics[width=\textwidth]{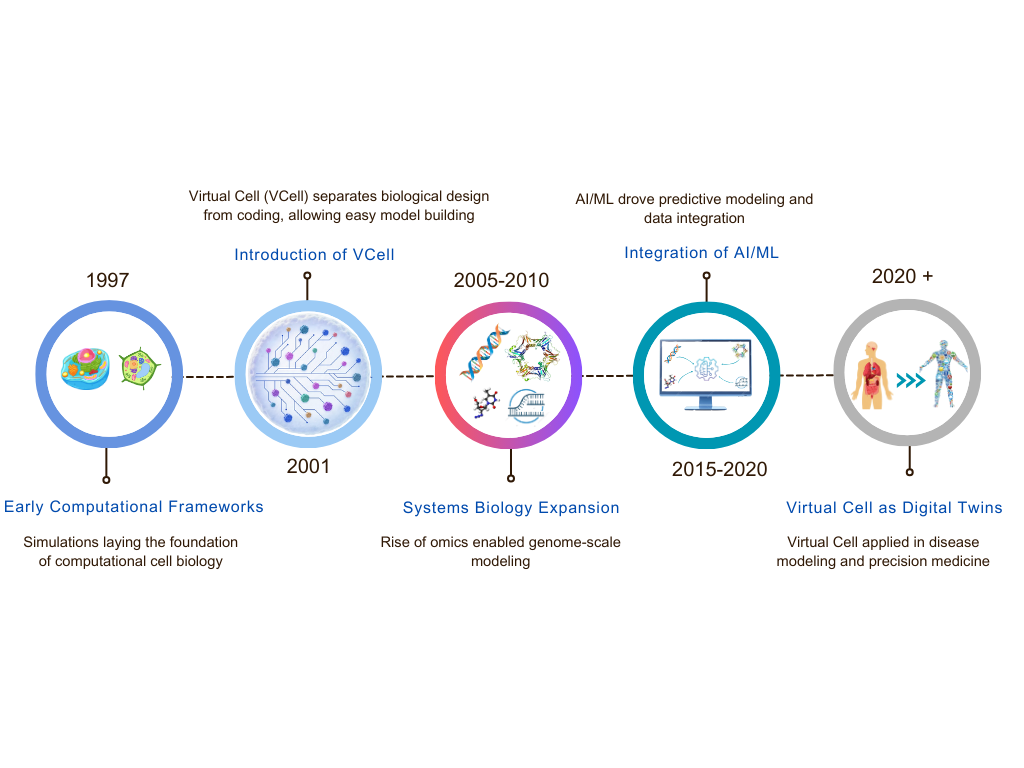}
  \caption{Evolution of the Virtual Cell from early computational frameworks to systems biology expansion, AI/ML integration, and digital twins}
  \label{fig:Evolution-virtual-wide}
\end{figure*}

In the last twenty years, the field has moved quickly thanks to improvements in genomics, proteomics, metabolomics, and high-resolution imaging. This has led to an unprecedented amount of cellular data \cite{Buenrostro2018Integrated},\cite{li2022fly}.  These datasets facilitated genome-scale metabolic models, dynamic signaling maps, and integrated whole-cell frameworks.  Recently, the integration of biological systems with artificial intelligence has enhanced the abilities of Virtual Cells, facilitating the management of, noisy high-dimensional data and enabling predictive modeling at previously unreachable scales \cite{li2022fly}. Unlike traditional modeling that rely heavily on well established equations, AI methodologies may discern patterns directly from data, facilitating the extraction of mechanistic insights from extensive biological datasets \cite{feldman2019optical},\cite{qi2013repurposing}.  Machine learning methodologies enable the discovery of biomarkers, prediction of pathway activity, and optimization of model parameters.  Deep learning architectures, such as convolutional and recurrent networks, have been used for imaging data and temporal omics datasets, effectively capturing spatial and dynamic characteristics of cellular systems \cite{qi2013repurposing},\cite{Porto2020BaseEditing}.
For multi-scale integration, AI plays a critical role by consolidating different datasets including DNA sequences, transcriptome profiles, protein networks, and cell imaging into cohesive Virtual Cell frameworks\cite{feldman2019optical}, \cite{Anzalone2020GenomeEditing}. Researchers may connect raw biological data with functional cell-level models by utilizing AI-driven feature extraction, dimensionality reduction, and predictive modeling. Moreover, reinforcement learning and Bayesian optimization techniques have been used to refine model performance, enhancing simulation precision and flexibility \cite{feldman2019optical},\cite{dixit2016perturb}.

\section{Vision and Principles of Virtual Cells}
The long-term vision of Virtual Cells is to create computationally designed analogs of living cells that function as predictive, testable, and scalable platforms for biological investigation. The initial investigation marks a notable progression, as it introduced the first detailed whole-cell computer model of Mycoplasma genitalium \cite{karr2012whole} . The conversion of genotypic data into phenotypic predictions is exemplified by the integration of 28 sub-models, which include processes like DNA replication, metabolism, and regulatory mechanisms. This proof-of-concept demonstrated that intricate biological processes can be integrated into a singular predictive model, serving as a fundamental principle of the Virtual Cell.
The earlier researchers highlighted inconsistencies across different data sources and demonstrated the potential of mechanistic simulations as comprehensive platforms \cite{macklin2020simultaneous}. They emphasized the significance of benchmarking and cross-validation as critical components, underscoring the necessity for Virtual Cell models to be both predictive and dependable, while integrating stringent quality control protocols.

\begin{figure*}[t]
  \centering
  \includegraphics[width=\textwidth]{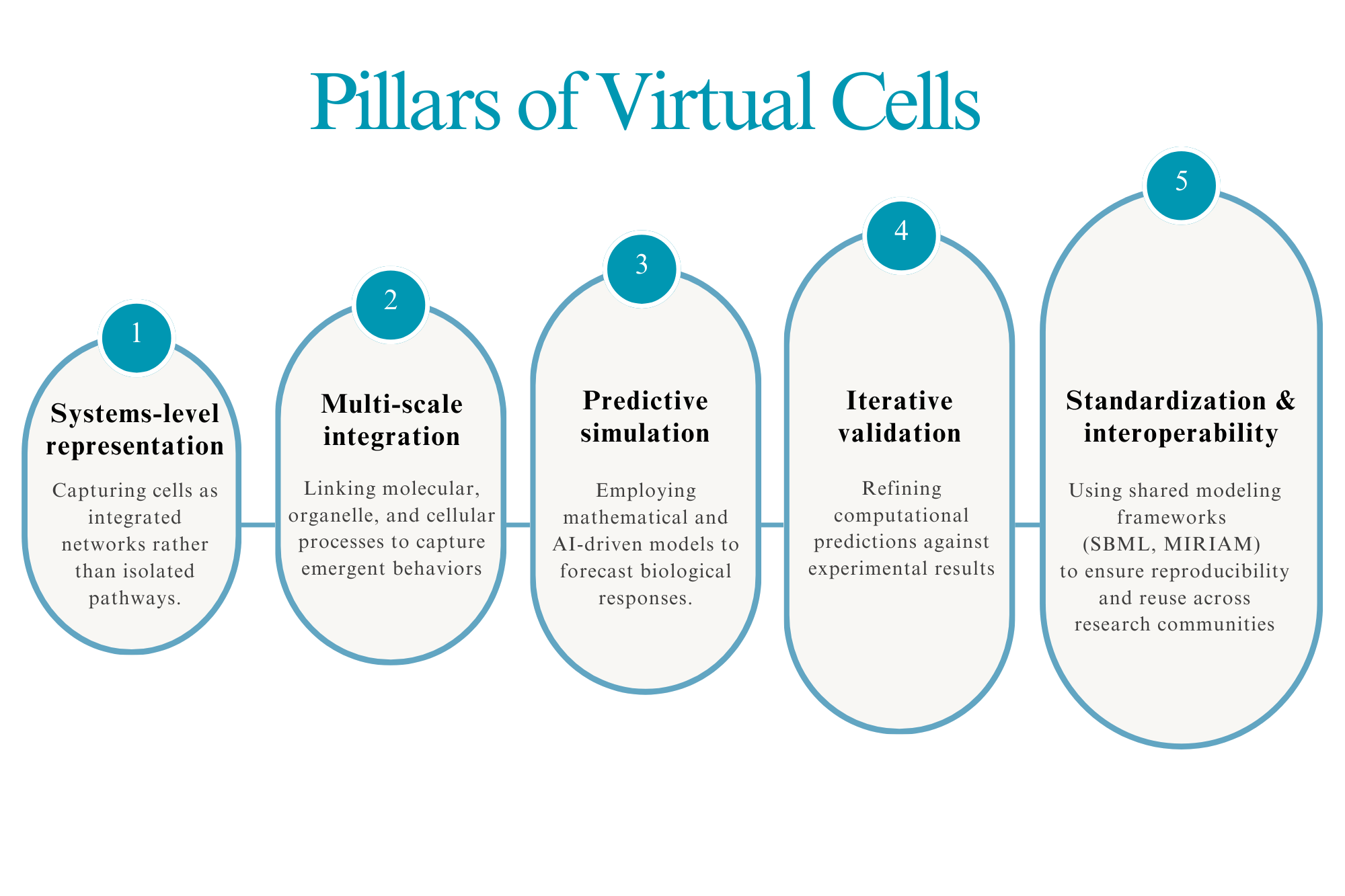}
  
  \caption{Five pillars of Virtual Cells: (1) systems-level modeling, (2) multi-scale integration, (3) predictive modeling, (4) Iterative validation, and (5) standardization, and interoperability.}
  \label{fig:Pillarsof virtual-cell-wide}
\end{figure*}

The evolution of the Virtual Cell is shown in Figure ~\ref{fig:Pillarsof virtual-cell-wide}. The concept has been significantly influenced by data-driven methodologies, in conjunction with mechanistic frameworks.  Yuan et al.\cite{yuan2021cellbox} introduced CellBox, which combines statistical inference, machine learning, and mechanistic biology to clarify reactions to perturbations. Although CellBox does not constitute a comprehensive Virtual Cell implementation, it illustrates the amalgamation of machine learning with mechanistic modeling, underscoring the effectiveness of hybrid methodologies for predictive simulations of cellular activity.

At the same time, comprehensive cell mapping efforts have provided the essential evidence required for the advancement of Virtual Cell technology. The Human Cell Atlas initiative introduced a comprehensive strategy to document all human cell types through the use of transcriptomics, epigenomics, and imaging techniques \cite{tanay2017scaling}   .     The Tabula and Sapiens initiatives advanced this viewpoint by offering multi-organ atlases that capture cellular diversity across tissues in both mice and humans \cite{the2022tabula}.The available resources ensure exceptional resolution.
Virtual cells differ from traditional static models by being designed as dynamic frameworks that continuously incorporate new data, allowing evolution in tandem with experimental insights \cite{singhvi1994engineering}, \cite{kalos2011t}.  This notion aligns with the primary goal of developing digital twins for biology, where computer simulations of cells enable therapeutic design, disease monitoring, and applications in synthetic biology \cite{czi2025cz}.   Virtual Cells function as dynamic intermediaries between data and experimentation, with the objective of reducing experimental costs, accelerating hypothesis development, and improving the translation of basic discoveries into therapeutic applications \cite{hie2019efficient}.

We may observe the Virtual Cells in five foundational pillers . The first is systems-level modeling, which depicts biological processes as interconnected networks encompassing metabolism, regulation, and signaling \cite{cong2013multiplex}.    The second is multi-scale integration, ensuring coherent connections among molecular, organelle, and cellular data to capture emergent events. The third is predictive modeling, which employs computational, statistical, and increasingly AI-driven methods to anticipate the outcomes of genetic modifications or medical treatments \cite{feldman2019optical},\cite{qi2013repurposing}.    
The fourth piller is Iterative validation is an essential concept that entails the systematic comparison of computational predictions with experimental data to enhance model accuracy and biological relevance \cite{michaelis1913kinetik}. Finally, the fifth pillar is the standardization and interoperability of Virtual Cells by employing frameworks like SBML and MIRIAM, which enable the reuse and enhancement of models across research groups \cite{li2022fly}.

The integration of these concepts forms a basis for the advancement of Virtual Cells as sophisticated tools in biomedicine, enabling rational drug design, precision diagnostics, and predictive synthetic biology \cite{brbic2022annotation},\cite{Bepler2021ProteinLanguage}.
Loew and Schaff first introduced the Virtual Cell as a problem-solving environment that bridges mathematical modeling and experimental cell biology while deliberately separating biological problem formulation from numerical implementation so that non-programmers can build models\cite{loew2001virtual}. VCell supports compartmental and fully spatial simulations via reaction-kinetic ODEs, diffusion/reaction-diffusion PDEs, and both deterministic and stochastic solvers. The use of geometry derived from microscopy, which allowed for realistic depictions of membranes, organelles, and entire cells, was a revolutionary development.  The work markedly improved accessibility, reproducibility, and spatial realism, with compelling demonstrations in electrophysiology, transport, and multiscale biochemical networks. A trade-off, however, is reduced solver efficiency in complex geometries and the “stair-step” artifacts that can arise from orthogonal meshing.  This study established the conceptual framework for virtual cell environments, focusing on modularity, interoperability, and usability.

Expanding the scope, Slepchenko et al. positioned VCell at the center of quantitative cell biology, particularly for spatially explicit intracellular models \cite{slepchenko2003quantitative}. Their review showed how spatial distributions shape transport, reaction kinetics, and signaling in ways that well-mixed models cannot capture. The key conclusion was that spatial modeling both explains observed behaviors and predicts emergent phenomena—gradients, oscillations, waves, and localized activation—while enabling iterative improvement with experimental feedback.
Resasco et al. offer a clear tour of VCell’s inner workings—from solver architecture to supported simulations \cite{resasco2012virtual}. The platform integrates finite-difference and finite-volume spatial solvers, electrophysiology modules, stochastic biochemical network simulators, and adaptive time stepping within a modular pipeline spanning image-based mesh generation through simulation and analysis. By tightly coupling modeling and imaging, VCell improves spatial fidelity for reaction–diffusion processes and electrophysiology. Strengths include flexibility across biological scales and the ability to treat deterministic and stochastic dynamics within one environment; principal constraints arise from computational cost at high spatial resolution and from managing large parameter spaces.
Herrington and Wang examine the issues related to clinical heterogeneity and the shortcomings of conventional biomarkers\cite{herrington2023clinical}. Despite significant progress in multi-omics profiling, the predictive power of individual molecular features or even combined signatures continues to be limited.  To tackle this issue, the authors outline innovative computational technique like manifold learning and Convex Analysis of Mixtures (CAM) that uncover latent structures within high-dimensional data.  Utilizing metabolomic and lipidomic datasets from extensive cohorts, these techniques uncovered consistent low-dimensional manifolds that illustrate fundamental biological processes.  The investigation highlights that these methodologies show potential for connecting molecular intricacies with insights that are pertinent to clinical applications.  Within the realm of Virtual Cells, these approaches demonstrate how sophisticated analytics can simplify complexity, reveal emerging patterns, and enhance the clarity of multi-omics models.
Moraru et al. described a database-driven, client–server architecture that enables collaborative modeling and scalable computation~\cite{moraru2008virtual}. Biological systems are specified, transformed into ODE/PDE/stochastic formalisms, and executed on distributed resources through a graphical interface. The workflow includes geometry import from imaging data, problem-type–aware solver selection, and spatial discretization. Benefits include integration of diverse processes on a single platform, model provenance and sharing, and efficient use of compute infrastructure.

Slepchenko and Loew further underscored the centrality of spatial modeling for discovering unexpected behaviors in cellular dynamics\cite{resasco2012virtual}. Using multi-compartment reaction–diffusion PDEs, hybrid ODE–PDE coupling, and parameter fitting to experimental data, they illustrated how organelle compartmentalization, molecular buffering, and diffusion barriers can qualitatively alter system behavior. Algorithms include stochastic simulators for low-copy-number species alongside deterministic solvers for transport-reaction systems. Advantages include mechanistic insight, predictive power, and linkage from molecular events to whole-cell outcomes; challenges include the heavy compute demands of 3D models and difficulties validating predictions where spatial measurements are limited.
An updated view emphasized hybrid, stochastic, and deterministic simulations within a unified framework, stronger spatial solvers, and a database backbone that streamlines collaborative workflows~\cite{heydari2017development}. The typical pipeline builds geometry from images, specifies reaction networks, and configures solvers for coupled transport and signaling. Implemented algorithms span stochastic kinetic schemes, membrane-transport models, and finite-difference/finite-volume PDE solvers. Benefits include seamless combination of modeling paradigms and applicability across scales; limitations include dependence on high-quality images for accurate geometry and substantial resources for large 3D problems~\cite{liu2025dvlo,jiang20243dsflabelling}.

Finally, Schaff et al. provided an early demonstration of nonlinear biochemical dynamics on image-derived geometries, coupling reaction networks with spatial diffusion to simulate pattern formation, wave propagation, and threshold responses \cite{schaff2012virtual}. The main finding was that geometry can qualitatively reshape dynamics, sometimes contradicting predictions from simpler, well-mixed models. The work combined stochastic kinetics with deterministic PDE solvers and used microscopy for validation, delivering both mechanistic insight and practical guidance for experiment. Remaining challenges include computational expense and sensitivity to parameter accuracy and mesh resolution.
Together, these contributions articulate a coherent vision: Virtual Cells merge image-informed spatial realism, hybrid mathematical formalisms, and collaborative, reproducible workflows to produce interpretable, predictive models that meaningfully guide biological discovery~\cite{moraru2008virtual}.


\section{Technological Foundations}
\begin{figure}[htbp]
  \centering
  \includegraphics[width=\textwidth]{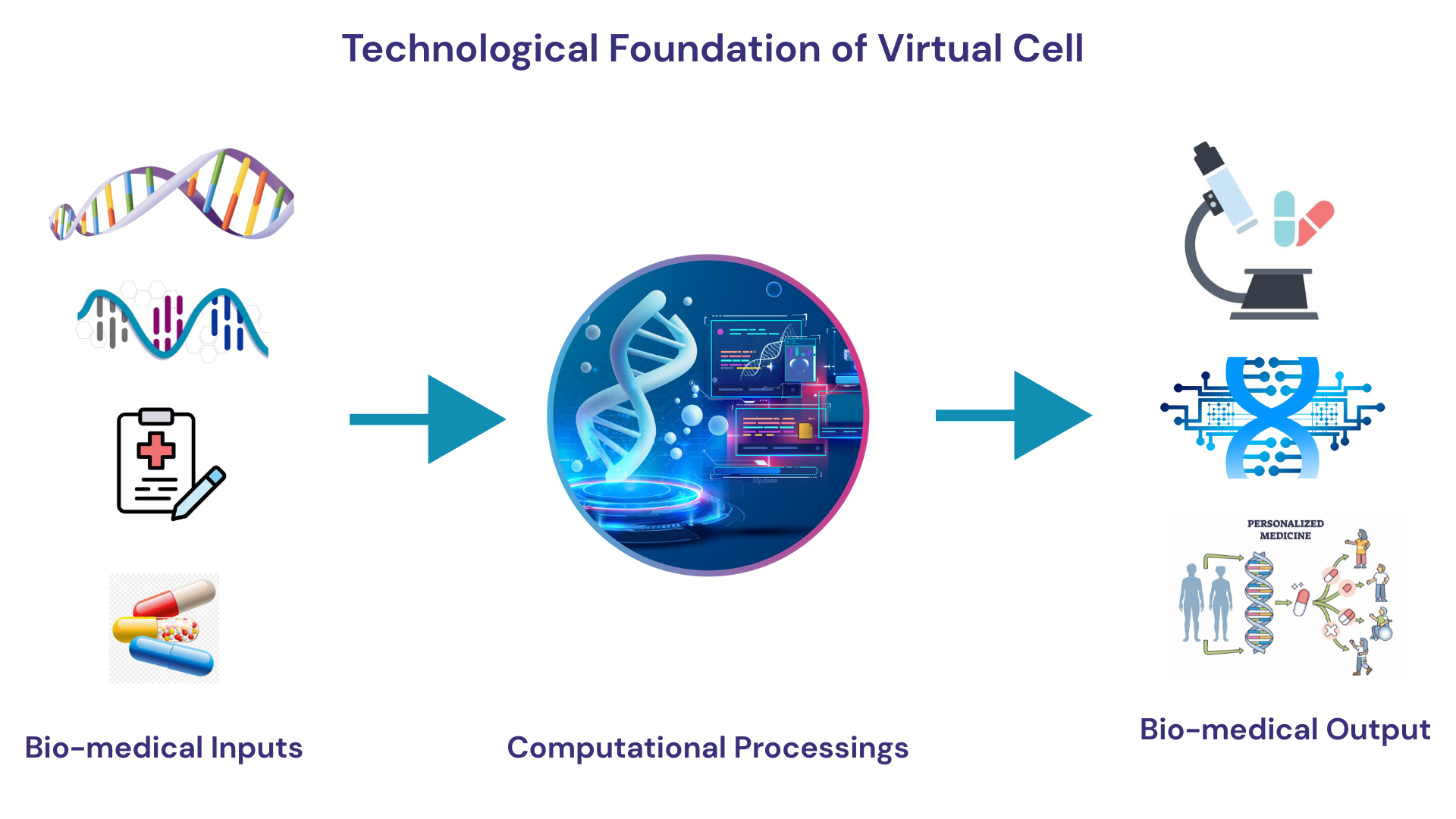}
  \caption{Technological foundation of Virtual Cell, illustrating how diverse biomedical inputs are processed computationally to generate predictive bio-medical outputs.}
  \label{fig:tech_foundation_virtual_cell}
\end{figure}

The Virtual Cell framework relies on a robust technological foundation that converts multidimensional biological data into valuable predictions.   The process starts with an acquisition of biomedical inputs, such as omics datasets, clinical data, and experimental assays, as shown in Figure ~\ref{fig:tech_foundation_virtual_cell}. The inputs are processed using computational methods to accurately capture how cells move and change.The biomedical outputs enable applications that include personalized medicine, drug discovery, and a comprehensive understanding of disease mechanisms at a systems level.  This foundation makes the Virtual Cell not just a place to run simulations, but also a tool that connects experimental biology with clinical translation. 
The Virtual Cell approach highlights the integration of multiple modalities to depict dynamic processes across molecular, spatial, and temporal dimensions~\cite{moraru2008virtual}.  Reviews have identified ongoing challenges in this integration, such as heterogeneity, noise, and scalability in the combination of transcriptomic, epigenomic, and proteomic data~\cite{argelaguet2021computational}.  Complementary benchmarking efforts and algorithmic advancements, such as Scanorama, illustrate the effective harmonization of large single-cell atlases, frequently without the need for explicit batch correction.  Ontology-driven resources such as Gene Ontology, KEGG, and Reactome~\cite{bard2005ontology,kanehisa2007kegg,fabregat2018reactome}  offer structured knowledge that aids in contextualizing integrated datasets.  These tools collectively form the foundation for effective data integration in Virtual Cell models.

Imaging and spatial biology complement omics by offering phenotypic and localization data that improve predictive modeling.  Standardized image-analysis platforms, such as CellProfiler 3.0~\cite{mcquin2018cellprofiler} and Cell Painting assays~\cite{bray2016cell}, in addition to spatial proteomics atlases~\cite{Thul2017SubcellularMap}, offer extensive morphological signatures and subcellular maps. Simultaneously, deep learning has facilitated scalable feature extraction and reproducibility in image-based analysis~\cite{moen2019deep}.  These advancements demonstrate the systematic integration of multimodal evidence into Virtual Cell pipelines.

Recent advancements in artificial intelligence have yielded effective techniques for the integration and interpretation of large-scale cellular data.  Frameworks like scVI~\cite{lopez2018deep} and scGen~\cite{lotfollahi2019scgen,lotfollahi2023predicting} employ probabilistic and representation-learning methods for tasks including dimensionality reduction, batch correction, and perturbation prediction.  In oncology, multiscale systems biology approaches combine molecular, cellular, and tissue-level data with deterministic, stochastic, and Bayesian models to forecast drug response and resistance mechanisms. Bayesian frameworks utilized in embryonic stem cell differentiation demonstrate that predictive modeling effectively captures abrupt lineage transitions, as validated by flow cytometry and live-cell imaging~\cite{jang2017dynamics}.  These studies demonstrate that AI/ML and multiscale modeling enhance the predictive capabilities of Virtual Cells. Yue Wang et al. provide a comprehensive examination of the multiscale modeling in endocrine-related cancers\cite{clarke2019systems}. The study emphasizes the integration of systems biology that amalgamates data from molecular, cellular, and tissue levels to elucidate the complexities of cancer.  The review presents an overview of several mathematical models, encompassing deterministic, stochastic, and Bayesian frameworks, as well as computational techniques for analyzing high-dimensional omics data. The authors notably tackle significant challenges such as parameter estimation, model robustness, and the potential for error propagation throughout workflows, while emphasizing the critical need for thorough validation.  The study illustrates the application of modular and multiscale frameworks to demonstrate how dynamic models can effectively predict cell fate decisions, drug responses, and mechanisms of resistance in ER+ breast cancer.  This study illustrates the technological and biomedical underpinnings of Virtual Cells, highlighting their capacity to enhance discovery and refine therapeutic approaches in oncology.\\
Jang et al. conduct a pivotal study that integrates single-cell transcriptomics with computational modeling to elucidate the dynamics of embryonic stem cell differentiation \cite{jang2017dynamics}.The authors utilized a Bayesian framework to delineate nine distinct cell states and their lineage transitions, demonstrating that differentiation transpires through abrupt, state-dependent modifications rather than gradual transformations.    A probabilistic gene regulatory network model was developed to demonstrate the varied responses of distinct cell states to modifications.     The investigation provided empirical validation for these predictions through the use of flow cytometry and live-cell imaging, thereby confirming the functional uniqueness of the inferred states.     These studies demonstrate that AI/ML and multiscale modeling enhance the predictive capabilities of Virtual Cells.

 Foundation models and transformer-based architectures enhance the scalability and generalizability of Virtual Cell frameworks.  Universal cell embeddings facilitate ~\cite{rosenuniversal,huang2025ai,huang2024ar}cross-dataset representations, whereas sequence-based deep learning methods like Enformer ~\cite{avsec2021effective} and HyenaDNA ~\cite{nguyen2023hyenadna} effectively capture long-range genomic dependencies.  These innovations enhance the predictability of genotype-to-phenotype relationships and establish AI as a key component in integrating omics, imaging, and spatial data into comprehensive predictive models. The advancements in the area of artificial intelligence make it a crucial element, enabling the integration of omics, imaging, and spatial data into comprehensive predictive models.
 
 This section outlines the three fundamental components of the Virtual Cell’s technological framework—multimodal modeling, data integration, and bioengineering, explains how they jointly empower predictive and translational applications. Multimodal modeling systematically encapsulates the heterogeneity of biological processes by integrating deterministic, stochastic, and hybrid models~\cite{du2024embracing}.Data integration utilizes advancements in multi-omics, imaging, and high-throughput technologies to ensure that models are based on empirical data. Bioengineering principles enable the transformation of computational models into scalable frameworks that inform experimental biology and biomedical applications.   The elements form the basis of the Virtual Cell, enabling researchers to link computational models with real biological systems.

\begin{table}[h!]
\centering
\caption{Comparative overview of the core technical foundations in Virtual Cell, summarizing key elements and their roles in enabling mechanistic modeling, integrative analysis, and translational bioengineering applications.}
\label{tab:technical_foundations}
\begin{tabular}{p{3cm} p{6cm} p{7cm}}
\hline
\textbf{Component} & \textbf{Key Elements} & \textbf{Role in Virtual Cell} \\
\hline
Multimodal Modelling & Deterministic models (ODE/PDE), stochastic simulations, hybrid approaches, and AI/ML-driven frameworks & Provides flexible strategies to capture biological complexity across molecular, cellular, and tissue scales, accommodating both predictable dynamics and inherent biological variability. \\
\hline
Data Integration & Multi-omics datasets (genomics, transcriptomics, proteomics, metabolomics), imaging technologies, and clinical records & Anchors computational models in diverse, high-dimensional empirical evidence, enabling holistic representations of cellular states and disease conditions. \\
\hline
Bioengineering & Synthetic biology tools, personalized medicine strategies, and computational drug design frameworks & Translates Virtual Cell models into practical biomedical applications, supporting rational therapy design, patient-specific predictions, and experimental validation. \\
\hline
\end{tabular}
\end{table}

\subsection{Multimodal modeling}
\begin{figure}[htbp] 
    \centering
    \includegraphics[width=0.7\textwidth]{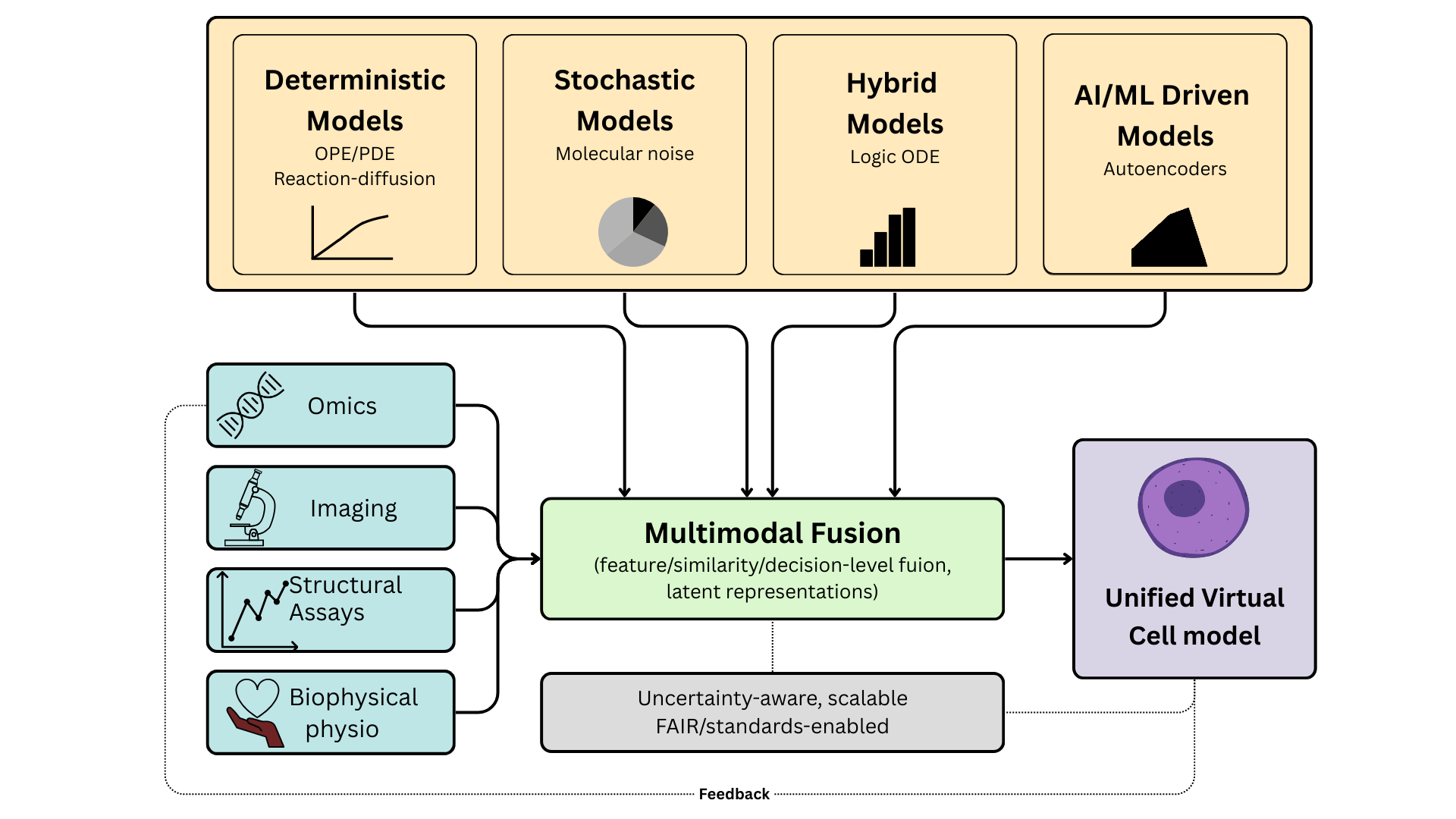} 
    \caption{Framework for Multimodal Fusion in Virtual Cell Modeling}
    \label{fig:multimodal-modeling-overview}
\end{figure}

Multimodal modeling is essential to the Virtual Cell framework, facilitating the integration of diverse methodologies and heterogeneous biological data to produce a comprehensive and verifiable representation of cellular function.   This strategy combines deterministic, stochastic, hybrid, and AI/ML-driven models utilizing diverse inputs such as omics, imaging, structural assays, and biophysical physiology.  Figure~\ref{fig:multimodal-modeling-overview} illustrates how these complementary inputs are integrated via multimodal fusion frameworks, which improve predictive accuracy, encompass various scales of biological complexity, and offer a scalable platform for simulating dynamic cellular behaviors.  This subsection delineates the historical development, methodological approaches, and integrative advancements that collectively characterize multimodal modeling in Virtual Cell research.

Initial Virtual Cell implementations paired deterministic solvers for reaction–diffusion PDEs with stochastic simulators to capture molecular noise~\cite{loew2001virtual}. Subsequent research built upon this foundation by integrating machine learning to incorporate various modalities into predictive models, highlighting the significance of multimodal fusion for the advancement of Virtual Cells.   Xu et al. demonstrated that representation learning, along with early, intermediate, and late fusion, enhances prediction and causal inference in the presence of noisy or incomplete modalities. They proposed taxonomies that differentiate between feature-, similarity-, and decision-level integration, while emphasizing the role of deep architectures, including autoencoders and variational schemes.In parallel, De Smet et al. systematized integration strategies, describing graph-based, kernel, and matrix factorization methods for aligning modalities, alongside Bayesian integration, network alignment, and manifold learning approaches~\cite{de2010advantages}. These taxonomies provide guidance on when each method is most beneficial for robust cellular simulations.

Advances in multimodal modeling have been particularly impactful at the single-cell level, where high-dimensional data streams must be harmonized. Stuart et al. devised methodologies for the integration of RNA, ATAC, protein, and spatial modalities, illustrating that manifold alignment, canonical correlation, and variational autoencoders improve cross-modal harmonization, despite persistent challenges such as over-correction and modality imbalance~\cite{stuart2019integrative}. Satija et al. illustrated that joint embeddings and graph neighborhoods successfully capture both shared and modality-specific signals, resulting in precise integrated atlases and perturbation models, while highlighting challenges including batch drift and missing features~\cite{stuart2019integrative}.  Johnson et al. defined "living" Virtual Cells that constantly combine simulations with volumetric imaging and knowledge graphs, making it easier to come up with new hypotheses and rebuild phenotypes.  But they face problems with accessing datasets, computing costs, and interoperability~\cite{johnson2023building}.   The contributions show that multimodal modeling makes cellular models more accurate, easier to understand, and more biologically realistic in a number of different situations.\\

At the mesoscale, Russel et al. illustrated the integration of cryo-EM, tomography, cross-linking, and proteomics with physics-based priors for the reconstruction of cellular structures. Probabilistic scoring enables the creation of uncertainty-aware models that outperform single-method assays~\cite{de2010advantages}.  Sali et al. conceptualized integrated structural biology as an optimization problem limited by several factors (EM, XL-MS, SAXS, FRET), demonstrating that Bayesian scoring and ensemble solutions yield dependable mesoscale assemblies for spatial Virtual Cell simulations~\cite{sali2010integrating}. Complementary work introduced scaffold abstractions—ontologically guided maps linking measurements to biological entities such as pathways, complexes, and compartments—that ease integration across assays and support hierarchical model construction~\cite{ideker2003building}. These structural approaches furnish the geometries and constraints necessary for linking multimodal data to spatially resolved models.\\
Multimodal modeling enables an integrated representation of processes across various scales, from molecular to organ levels.   Viceconti et al. delineated the principles of weak and strong coupling, PDE/ODE co-simulation, and surrogate modeling within the framework of the Virtual Physiological Human initiative, presenting case studies in cardiac electrophysiology and metabolism~\cite{viceconti2008virtual}. These strategies minimize error propagation across scales and facilitate modular calibration and validation.  Karr et al. presented a significant example through their extensive model of Mycoplasma genitalium, which incorporated 28 submodels covering gene expression, metabolism, and regulation to replicate phenotypes and produce testable predictions~\cite{karr2012whole}. Together, these approaches demonstrate how multimodal modeling connects molecular, cellular, and tissue-level processes into predictive whole-cell frameworks.\\

Finally, the quantitative framework introduced by Slepchenko et al. linked physical laws to microscopy-derived geometries through spatially resolved reaction–diffusion PDEs, compartmental ODEs, stochastic solvers, and parameter estimation~\cite{slepchenko2003quantitative}. Spatial models demonstrate superior performance compared to well-mixed assumptions in processes like calcium signaling and nucleocytoplasmic transport, thereby enhancing hypothesis testing, experimental design, and educational efforts in cell biology.  This study established a foundation for integrating multimodal data into predictive Virtual Cell models, despite challenges related to solver performance, parameter identifiability, and scaling to complex geometries.Multimodal modeling establishes a framework enabling Virtual Cells to integrate diverse measurements, analyze biological mechanisms and scales, and produce empirically testable mechanistic predictions.  Integrating omics, imaging, structural, and physiological data within hybrid computational frameworks enhances the Virtual Cell, making it a robust tool for discovery and translational applications.\\
\begin{table}[htbp]
\centering
\caption{Comparison of modeling approaches applied in Virtual Cell frameworks. The table contrasts deterministic, stochastic, hybrid, and AI/ML-based methods by summarizing their defining characteristics, strengths in handling biological systems, and practical limitations, highlighting their complementary roles in advancing mechanistic and data-driven cell modeling.}
\label{tab:modeling_approaches}
\renewcommand{\arraystretch}{1.3}
\begin{tabularx}{\textwidth}{p{3cm} X X X}
\hline
\textbf{\textcolor{black}{Modeling Approach}} & \textbf{Characteristics} & \textbf{Advantages} & \textbf{Limitations} \\
\hline
Deterministic (ODE/PDE) & Relies on continuous, equation-based formulations such as ordinary and partial differential equations to describe system dynamics under defined initial conditions. & Provides high precision in well-controlled and predictable systems, allowing detailed simulations of biochemical kinetics and metabolic pathways. & Struggles to capture biological noise, stochasticity, and heterogeneity inherent in real cellular systems. \\
\hline
Stochastic & Uses probability-driven frameworks to represent random fluctuations in molecular interactions and gene expression. & Accurately models noise, variability, and rare events in biological processes, reflecting single-cell diversity and randomness in gene regulation. & Often computationally intensive; scaling becomes difficult for large networks or multi-scale simulations. \\
\hline
Hybrid & Integrates deterministic and stochastic elements, applying continuous equations to stable processes and random modeling to inherently variable processes. & Provides a balanced framework that combines precision with realism, making it suitable for complex biological systems where both order and randomness coexist. & More difficult to implement and calibrate; requires careful partitioning of processes into deterministic vs. stochastic domains. \\
\hline
AI/ML-based & Employs machine learning and artificial intelligence models to learn patterns directly from high-dimensional biological datasets. & Scales efficiently to massive datasets, is adaptive to diverse biological contexts, and is capable of uncovering hidden nonlinear relationships. & Dependent on quality and representativeness of training data; often considered a ``black-box'' with limited interpretability. \\
\hline
\end{tabularx}
\end{table}

\subsection{Data Integration}
Data integration is essential to Virtual Cell modeling, enabling the conversion of various biological information into unified and mechanistically relevant representations.   Recent advancements in single-cell RNA sequencing (scRNA-seq), spatial transcriptomics, and high-content imaging have generated comprehensive, multidimensional datasets that precisely depict biological states at the single-cell level, while preserving spatial and morphological context. These inputs impose constraints that bulk assays fail to capture; however, they also present challenges including noise, batch effects, and platform variability.  Effective Virtual Cell modeling therefore depends on robust strategies for ingesting, harmonizing, and validating these data to yield interpretable parameters and geometries that can map back to experimental biology.

Within VCell environments, dedicated ingestion pipelines can directly process imaging and omics modalities, linking experimental observations with model parameterization and boundary conditions~\cite{schaff2016rule}. Integration generally proceeds in a closed loop, where initial experimental data seed model structures, simulations are validated against new measurements, and discrepancies drive targeted updates. Machine learning has accelerated this process by introducing principled fusion strategies. Representation-level fusion—learning shared latent spaces via autoencoders or variational autoencoders—has proven especially effective in improving calibration and robustness compared to single-modality pipelines. Methodologically, the field now distinguishes early/feature fusion, intermediate/representation fusion, and late/decision fusion, each with specific trade-offs in uncertainty handling and interpretability.

Graph-centric and kernel methods provide a complementary pathway for integrating diverse assays. By embedding omics, imaging, and clinical data into shared similarity or network spaces, these methods expose functional modules and cross-modal correspondences that would be invisible in isolated datasets~\cite{loew2001virtual}. Toolkits such as graph kernels, probabilistic network alignment, matrix factorization, and multi-view learning support integration while tolerating missing or noisy inputs. However, their utility depends on careful treatment of heterogeneous noise models and cross-modal validation to avoid spurious associations.

At the mesoscale, structural biology methods have been adapted for integrative modeling. Techniques such as cryo-EM, tomography, XL-MS, SAXS, and FRET can be combined with probabilistic scoring to produce ensemble reconstructions that quantify uncertainty and outperform single-assay models. Frameworks such as IMP and TEMPy utilize Bayesian or maximum-entropy scoring, extensive sampling, and cross-validation with withheld restraints to produce geometries appropriate for reaction–diffusion solvers in VCell simulations.  Despite their strengths, these methods encounter limitations such as uneven data coverage, inconsistent error models, and significant computational costs, highlighting the necessity for standardized repositories and community benchmarks.Automating the assembly of biochemical networks from heterogeneous sources accelerates model drafting, but simulation-grade models still depend on expert curation, conflict resolution, and context specificity \cite{walker2009virtual}. Practical pipelines couple ontology alignment with motif discovery, topology inference, parameter estimation, and constraint checks, producing draft signaling and metabolic maps that are iteratively refined against experiments. The main risks are identifier inconsistencies and error propagation from upstream annotations, reinforcing the value of validation-driven refinement before handing networks to mechanistic solvers.

Predictive modeling in Virtual Cells requires linking signaling, transcriptional regulation, and metabolism rather than treating them in isolation. Multi-layer designs capture cross-layer feedbacks that reshape dynamics and drug response~\cite{chen2024applying}. These approaches enable analyses of synergy prediction and metabolic rewiring, though they also introduce computational stiffness, parameter identifiability issues, and high data demands. Standards-aware composition is therefore essential to allow knowledge to flow consistently across layers within VCell frameworks.

At the single-cell level, careful handling of batch effects and cross-platform alignment remains central. Various methods such as canonical correlation, mutual nearest neighbors, variational autoencoders, and optimal transport have been employed to integrate RNA, ATAC, protein, and spatial measurements, thereby aiding in the creation of atlases, lineage inference, and the initiation of simulations~\cite{chen2021data}.   Manifold and graph-based alignment improves cross-modal harmonization; however, excessive correction can obscure genuine biological signals.   Joint embeddings and graph neighborhoods, as demonstrated by Stuart and Satija, effectively capture both shared and modality-specific signals, thereby improving reference mapping and perturbation analysis~\cite{stuart2019integrative}.  Despite of these advancements, the challenges of scaling to millions of cells and addressing modality imbalance persist. This situation encourages the implementation of uncertainty-aware mappings and transfer learning to enhance generalizability across various laboratories and tissues.

Whole-cell models illustrate the end-to-end impact of integration. The Mycoplasma genitalium model integrates 25–30 submodels that encompass gene expression, metabolism, and regulation, facilitating the replication of growth phenotypes and the formulation of testable hypotheses~\cite{karr2012whole}.  Scaffold-based frameworks have been proposed to connect heterogeneous measurements to biological entities, including pathways, complexes, and compartments, facilitating hierarchical model construction and simulation initialization~\cite{ideker2003building}.  These designs are limited by incomplete ontologies and the necessity for manual curation, yet they demonstrate how structured mappings enable the conversion of raw data into frameworks suitable for simulation.

Finally, quantitative integration closes the loop between data and simulation. The VCell environment has been developed to pair imaging-derived geometries with mechanistic solvers—reaction–diffusion PDEs, compartmental ODEs, and stochastic methods—under database-backed provenance~\cite{slepchenko2003quantitative}. Spatially resolved models consistently outperform well-mixed assumptions for processes such as nucleocytoplasmic flux and calcium signaling, providing a platform for hypothesis testing, experimental design, and education. Remaining limitations, such as solver cost, parameter identifiability, and restrictions on modeling multiple organelles simultaneously, reinforce why rigorous data integration is not optional but foundational for Virtual Cell modeling. Data integration makes sure that Virtual Cells go from being just abstract simulations to being real, predictive frameworks.  Data integration is the key link that connects experimental measurements to mechanistic cellular models by bringing together omics, imaging, structural, and clinical data in scalable computational pipelines. This makes both discovery-driven research and translational applications possible.

\begin{figure}[htbp] 
    \centering
    \includegraphics[width=0.7\textwidth]{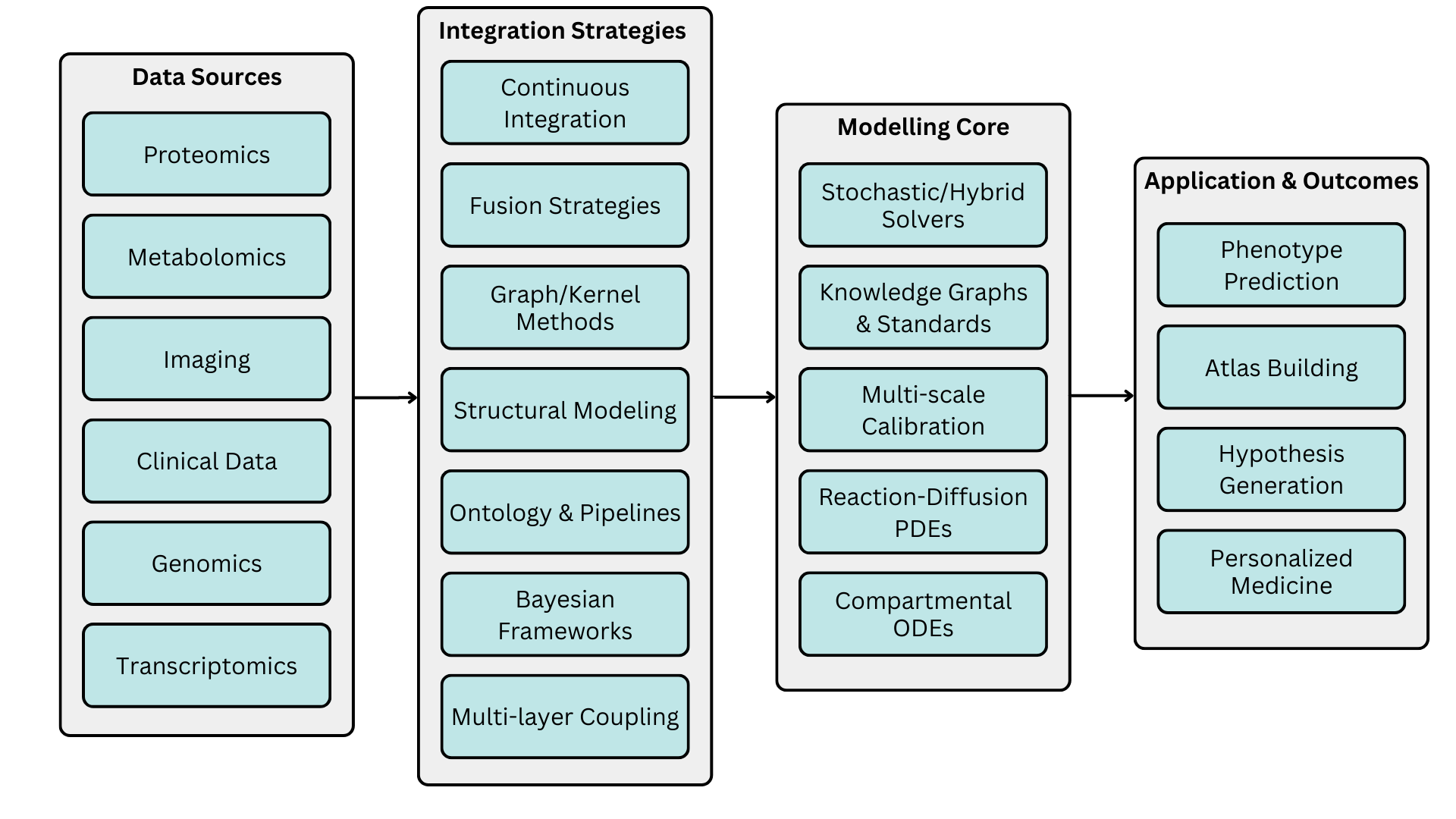} 
    \caption{Framework for Data Integration in Virtual Cell Modeling: from multimodal data sources to applications and outcomes}
    \label{Strategies}
\end{figure}

\begin{table}[htbp]
\centering
\caption{Overview of data integration modalities in Virtual Cell. The table outlines diverse data sources (genomics, transcriptomics, proteomics, metabolomics, imaging, and clinical records), the technologies used to generate them, and the corresponding integration strategies. Together, these approaches provide the empirical foundation for constructing robust, multimodal Virtual Cell models capable of linking molecular events with cellular and clinical phenotypes.}
\label{tab:data_integration_modalities}
\renewcommand{\arraystretch}{1.25}
\begin{tabularx}{\textwidth}{p{3.1cm} X X}
\toprule
\textbf{Data Source} & \textbf{Example Technologies} & \textbf{Integration Strategy} \\
\midrule
Genomics &
High-throughput DNA sequencing; whole-genome sequencing (WGS); genome-wide association studies (GWAS); CRISPR-based functional screens &
Network inference linking variants to pathways; construction of gene-regulatory networks to predict interactions and disease susceptibility \\
\addlinespace
Transcriptomics &
Bulk RNA-Seq; single-cell RNA-Seq (scRNA-Seq); spatial transcriptomics; microarrays &
Co-expression network building and dynamic transcriptional maps; clustering to identify cell states and tissue-specific programs \\
\addlinespace
Proteomics &
Mass spectrometry (MS), tandem MS/MS; protein microarrays; phosphoproteomics &
Integration via protein–protein interaction (PPI) maps, signaling-network reconstruction, and pathway enrichment analyses \\
\addlinespace
Metabolomics &
Liquid chromatography–mass spectrometry (LC–MS); gas chromatography–mass spectrometry (GC–MS); nuclear magnetic resonance (NMR) spectroscopy &
Mapping to metabolic pathways; flux-balance analysis (FBA); integration with proteomics for metabolic network modeling \\
\addlinespace
Imaging &
Confocal and multiphoton microscopy; fluorescence-lifetime imaging (FLIM); live-cell super-resolution imaging &
Spatial–temporal mapping of molecular processes; incorporation into spatially resolved computational models of cellular dynamics \\
\addlinespace
Clinical Data &
Electronic health records (EHRs); patient cohorts; clinical-trial repositories; wearable/remote monitoring devices &
Knowledge graphs and machine-learning frameworks for multi-modal fusion; linking clinical phenotypes to molecular data to enable personalized medicine \\
\bottomrule
\end{tabularx}
\end{table}

\subsection{Bioengineering}
Advancements in bioengineering have played a crucial role in converting Virtual Cells from theoretical constructs into effective instruments for simulating and designing cellular functions. By embedding principles of modularity, multi-scale integration, and reproducibility, computational frameworks now support applications ranging from synthetic biology to therapeutic development. Goldberg et al. demonstrated that whole-cell modeling is achievable when constructed from modular, testable submodels that can be integrated into extensive simulations.   The importance of hybrid deterministic–stochastic schemes, constraint-based layers, and Bayesian calibration in improving design–build–test–learn (DBTL) loops in synthetic biology was highlighted, along with the challenges associated with parameter identifiability and computational cost.  Loew and Schaff described VCell as a database-driven, client–server environment that converts microscopy-derived geometries into reaction–diffusion PDEs, compartmental ODEs, and stochastic processes~\cite{loew2001virtual}.  This framework ensured reproducibility and traceability; however, challenges related to meshing fidelity and solver performance were acknowledged.\\

Rekhi and Qutub proposed methods in mammalian systems that integrate top-down inference with bottom-up module engineering.  The methodology incorporated ODE/PDE dynamics, constraint-based flux models, and statistical learning to optimize therapeutic circuits and tissue-level behaviors.  Walker and Southgate put forward a "middle-out" paradigm, conceptualizing virtual cells as intermediaries that connect bottom-up molecular specifics and top-down tissue limitations~\cite{walker2009virtual}. By combining agent-based layers with ODE/PDE biochemistry, they addressed heterogeneity and microenvironmental inputs, though scaling and computational costs remain limiting.\\

Bioengineering frameworks have been extended to accelerate drug discovery. Chen et al. integrated spatiotemporal modeling, live-cell imaging, and single-cell trajectories to improve PK/PD translation and mechanism inference~\cite{chen2024applying}. Their toolset combined AI-based tracking, mechanochemical coupling, and diffusion–advection PDEs, supporting applications such as screen triage and rational drug combinations.\\

Resasco et al. detailed improvements to the VCell platform, encompassing spatial solvers, electrophysiology modules, stochastic simulators, and rule-based extensions~\cite{resasco2012virtual}.  Finite-volume discretization, adaptive time-stepping, and geometry import from microscopy facilitated multiscale modeling of Ca\textsuperscript{2+} dynamics and signal propagation.  Complementary studies on spatial stochastic simulations~\cite{nguyen2024absence} and mechanochemical coupling  emphasize the interplay of space, noise, and mechanics in influencing cellular decisions, while also indicating the necessity for smaller imaging-anchored models to enhance clarity.\\

Utilizing frameworks that combine dynamical VAEs, optimal transport, and state-space models to reconstruct lineages and simulate differentiation or therapeutic responses under uncertainty, time-series modeling has emerged as a prominent field of study. ~\cite{jung2025advances}. Control-theoretic approaches have shown that feedback architectures enhance robustness and tunability in engineered circuits, as formalized through nonlinear ordinary differential equations (ODEs) along with bifurcation and sensitivity analyses~\cite{you2004toward}.\\

Im et al. conducted a survey on strategies that connect molecular and cellular scales via coarse-grained dynamics, network kinetics, and continuum mechanics, highlighting the significance of cross-validation across these scales.  Foundational VCell studies have shown that spatially resolved reaction–diffusion partial differential equations, in conjunction with ordinary differential equations and stochastic solvers, more accurately replicate processes such as nucleocytoplasmic transport and Ca\textsuperscript{2+} signaling compared to well-mixed assumptions~\cite{slepchenko2003quantitative}.  Moraru et al. advanced this by creating GUI workflows, enabling distributed execution and collaborative sharing, which enhanced accessibility while also revealing performance limitations for complex geometries~\cite{moraru2008virtual}.Kaizu and Takahashi conducted a review of technologies that facilitate in silico whole-cell approaches at the ecosystem level, focusing on automation, multi-compartment simulations, and data governance~\cite{Kaizu2023}.  Benchmarking, shared pipelines, and standardization were advocated to tackle data harmonization and scalability issues.  Takahashi et al. contended that cellular simulations must incorporate principles from software engineering and configuration management to guarantee reproducibility and adaptability as models progress~\cite{takahashi2003cell}.Scaffold-based integration frameworks have improved the linking of datasets to pathways, complexes, and compartments, which has made hierarchical model composition stronger. However, they also highlight persistent issues, including incomplete ontologies and the need for manual curation~\cite{ideker2003building}.  The bioengineering innovations, including modular frameworks, spatiotemporal modeling, standards-aware platforms, and ecosystem coordination, convert Virtual Cells from descriptive models into predictive and design-oriented tools.  They facilitate both the generation of hypotheses and the application of findings in synthetic biology, drug discovery, and precision medicine.
\begin{figure}[htbp] 
    \centering
    \includegraphics[width=0.7\textwidth]{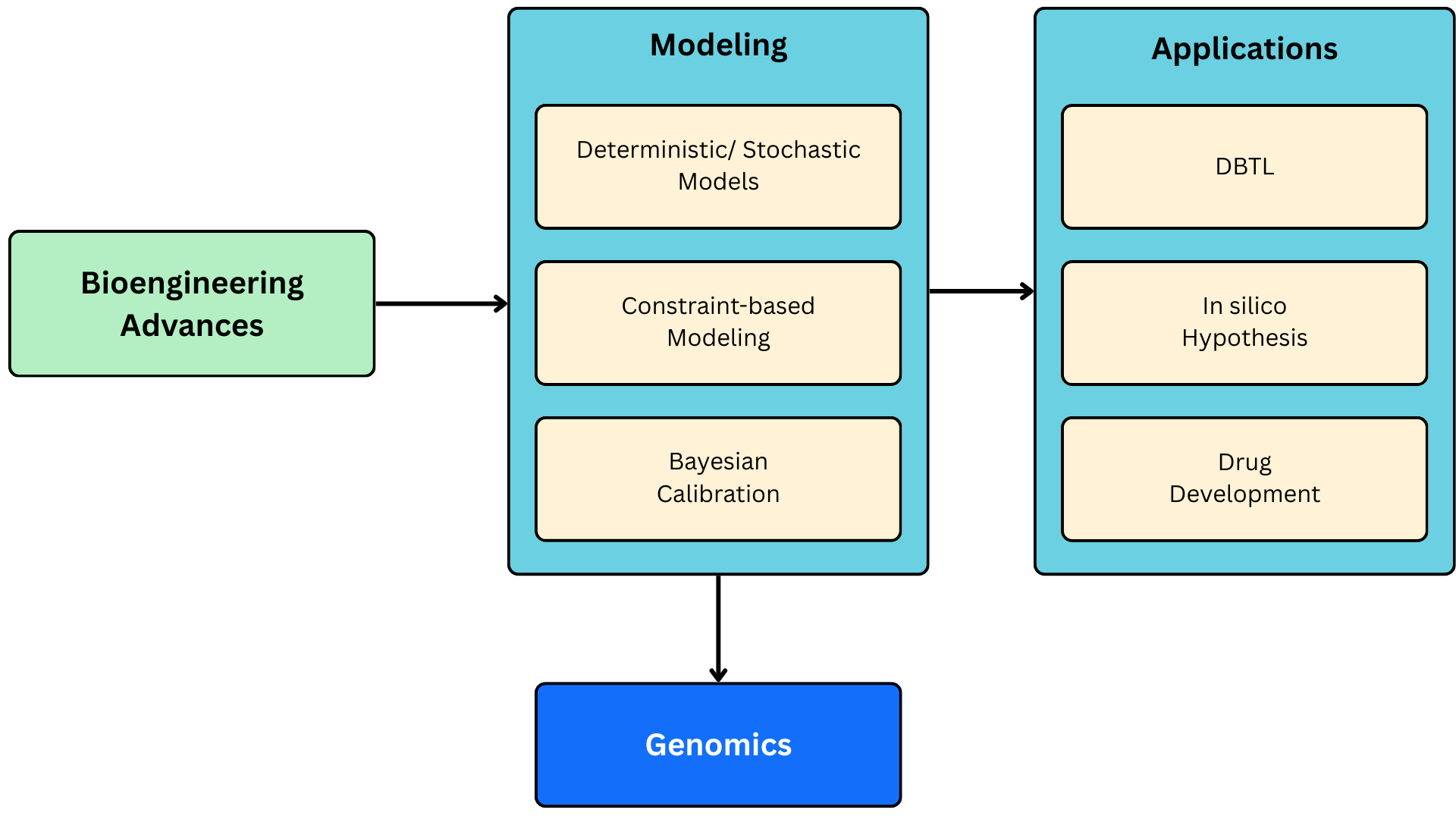} 
    \caption{Conceptual workflow illustrating the role of bioengineering innovations in modeling and their translation to genomics and biomedical applications.}
    \label{applications}
\end{figure}

\section{Benchmarking and Evaluation}

    The Virtual Cell Challenge has become a key community platform, giving us standard benchmarks to use when judging how well something can predict and how well it can be generalized across different biological settings.  By setting clear goals and metrics, curating reliable datasets, and requiring open reporting, it meets the field's urgent need for evaluation frameworks that can be reproduced and compared.  Recent versions have put more emphasis on end-to-end testing in real-world situations, making the results more reliable, easier to understand, and useful in the real world.  The platform codifies quantitative performance criteria, tracks the provenance of the data, and invites open participation. This creates a shared evaluative common that speeds up model development, thorough validation, and fair comparison.Zitnik et al.\ evaluate machine-learning methods for fusing heterogeneous biomedical data (omics, imaging, clinical/EHR) so that predictive models can be fairly compared between settings \cite{zitnik2019machine}. A central conclusion is that integration is not optional for reliable benchmarking: single-modality evaluations inflate performance and obscure failure modes. Methodologies are organized into early/feature fusion, intermediate/representation fusion (autoencoders/VAEs), and late/decision fusion, with attention to domain-shift diagnostics and provenance tracking. The objective is a unified taxonomy and best-practice playbook for multimodal benchmarking pipelines. The main applications are disease subtyping, pathway activity inference, and multimodal biomarker discovery, supported by practical dataset curation checklists. Recognized challenges, batch effects, limited labels, and cross-site drift, motivate standardized benchmarks and reports so that integrated outputs can be evaluated and reused in Virtual Cell settings.\\
    Gligorijević and Pržulj grounded cross-study and cross-species comparisons in graph-centric integration \cite{gligorijevic2015methods}. Network embeddings unify disparate assays into a shared space where link prediction and module recovery become benchmarkable tasks. Diffusion kernels and manifold learning expose conserved functional modules often missed by single modalities. The methodological suite—matrix factorization, graph kernels, probabilistic alignment, and multi-view learning—supports robust, scalable benchmarks while highlighting pitfalls such as inconsistent noise models and missing-not-at-random bias~\cite{perez2025fast}. They advocate leaderboards modeled after the Virtual Cell Challenge, augmented with probabilistic calibration and cross-modal validation.At the mesoscale, integrative structural modeling has become a benchmark domain. Joseph et al. demonstrated that hybrid restraints, including cryo-EM, tomography, XL-MS, SAXS/FRET, and proteomics, produce ensemble models accompanied by explicit confidence measures~\cite{fang2025respf}.  Cross-validated scoring utilizing withheld restraints demonstrates superior performance compared to individual techniques. This methodology is underpinned by Bayesian or maximum-entropy scoring, systematic sampling, and standardized evaluation splits, which facilitate equitable comparisons.   Alber et al. presented integrative structural biology as a form of probabilistic optimization through repository-oriented workflows (e.g., IMP), showing that ensemble models that include explicit uncertainty outperform single-method reconstructions~\cite{Sun2024RadiologyLLM,10800533,10957983}.   Thijssen et al. enhanced this framework by introducing Bayesian data-integration benchmarks that assess the contribution of each modality to performance~\cite{thijssen2018bayesian}.   Hierarchical models that integrate sparsity priors and posterior variance attribution improve calibration and interpretability relative to heuristic weighting. These studies collectively emphasize the advantages of uncertainty-aware benchmarking while also addressing the challenges associated with heterogeneous error models and significant computational costs.\\
    Single-cell resolution benchmarking has been thoroughly investigated. Luecken et al. evaluated atlas-level integration strategies, establishing benchmarks for scalability, batch correction, and the preservation of biological signals~\cite{lotfollahi2022mapping}.        Argelaguet et al. conducted an examination of matched and unmatched modalities, highlighting that excessive correction may obscure genuine signals, while insufficient correction could amplify false clusters.   The assessments included CCA, MNN, VAEs, and optimal transport, following established protocols and providing guidance for atlas-scale benchmarks.       Stuart and Satija introduced joint embeddings and graph neighborhoods for integrative single-cell analysis, showing improved downstream reproducibility across laboratories~\cite{stuart2019integrative}.   Efremova and Teichmann improved this study by evaluating cross-modal designs through the application of task-specific metrics and uncertainty-aware scoring~\cite{efremova2020computational}. Todorov and Saeys evaluated high-throughput single-cell pipelines, highlighting the necessity for metrics that maintain rare populations and standardized reporting practices~\cite{todorov2019computational}.  These efforts highlight the necessity for benchmarks to achieve a balance between correction and biological accuracy, accommodate large cohorts, and integrate uncertainty-aware evaluation.\\
    Whole-cell benchmarks offer predictive assessments that directly guide the development of Virtual Cell. Karr et al. developed predictive baselines using Mycoplasma genitalium, incorporating submodels aimed at replicating phenotypes and generating falsifiable predictions~\cite{karr2012whole}.     Benchmarks included growth phenotypes, knockouts, and sensitivity analyses, all validated by literature and experimental methods.     Ideker and Lauffenburger proposed scaffold-based integration to create benchmarks within biologically interpretable abstractions, such as pathways, complexes, and compartments~\cite{ideker2003building}.     Nilsson and Nielsen emphasized the significance of genome-scale metabolic reconstructions in cancer as essential benchmarks for detecting pathway reprogramming and therapeutic vulnerabilities~\cite{Anzalone2020GenomeEditing}, highlighting the need for validation via targeted assays and alignment with clinical practices.

     Viceconti et al. established benchmarking principles for the Virtual Physiological Human, emphasizing the necessity for tests that link molecular processes to organ-level dynamics~\cite{viceconti2008virtual}. Weak/strong coupling, surrogate models, and PDE/ODE co-simulation were proposed as methods to stabilize evaluations across scales. Verification and validation protocols reduce the risk of error propagation, but the absence of gold standards remains a limitation. Slepchenko et al. demonstrate that Virtual Cell benchmarking can be grounded in imaging-derived geometries and mechanistic solvers~\cite{slepchenko2003quantitative,zhou2025rho}.  Reaction–diffusion and compartmental models demonstrate superior performance compared to well-mixed assumptions, yielding consistent and testable results for spatial signaling and transport.  The aforementioned strands establish a benchmarking ethos for Virtual Cells characterized by uncertainty awareness, integration prioritization, cross-scale considerations, and adherence to community standards.  Future advancements will depend on increased participation in community challenges, the explicit adoption of FAIR principles, and the integration of AI-driven benchmarks, such as transformer-based foundation models, which will enhance the significance of benchmarking as a cultural norm in Virtual Cell development.

\section{Biomedical Applications}

    Virtual Cells are revolutionizing biomedical research by offering computational frameworks that simulate cellular processes with exceptional resolution.  Through the integration of genetic sequences, protein interaction networks, metabolic pathways, and imaging-derived geometries, dynamic environments for mechanistic simulations are established.  This integration allows researchers to analyze cellular mechanisms comprehensively, producing predictions that are both mechanistically sound and amenable to experimental validation (Wang et al., 2025).  These capabilities signify a transition from conventional reductionist methods to systems-level analyses essential for tackling multifactorial diseases and intricate therapeutic challenges.\\
    Virtual Cells serve as economical, highly efficient systems capable of evaluating thousands of potential compounds for safety, efficacy, and mechanisms of action prior to clinical trials.   This enhances the preliminary assessment procedures, diminishes reliance on animal models, and promotes the swift identification of off-target effects.    In personalized medicine, Virtual Cells can be adapted to mirror an individual's genetic and phenotypic characteristics, facilitating the development of customized therapeutic approaches that increase efficacy, reduce side effects, and improve disease prediction.  Additionally, they serve as computational laboratories for hypothesis-driven inquiries, allowing for the exploration of signaling pathways, metabolic fluxes, and microenvironmental dynamics without the costs or time commitments typically linked to large-scale experiments.   The applications illustrate how Virtual Cells function as tools that promote discovery and support evidence-based decision-making in the field of biomedical sciences.

\subsection{Drug Discovery}

Drug discovery is a significant domain for Virtual Cell applications, where computational platforms facilitate target identification and improve preclinical evaluation processes.   The Virtual Cell Challenge, detailed by Fahsbender et al. played a pivotal role in this initiative by developing extensive perturbation-response benchmarks that evaluate cross-context generalization, including variations across cell lines, doses, and time points\cite{fahsbender2025benchmarking}.   Their findings emphasize that methods validated only on restricted datasets often fail to perform satisfactorily when applied to diverse biological systems, thus highlighting the importance of context-sensitive modeling and thorough uncertainty assessment.They provides standardized benchmarks for evaluating predictive models in the context of realistic distributional shifts.   The organization of multimodal datasets and the establishment of reproducible tasks reduce the risks of overfitting and highlight models that show generalization across various cell types and experimental contexts.   The dynamic leaderboards enable continuous improvement and clarity.\\
Chen Li et al. played a pivotal role in this initiative by developing extensive perturbation-response benchmarks that evaluate cross-context generalization, including variations across cell lines, doses, and time points \cite{li2024benchmarking}.   Their findings emphasize that methods validated only on restricted datasets often fail to perform satisfactorily when applied to diverse biological systems, thus highlighting the importance of context-sensitive modeling and thorough uncertainty assessment. Wu et al. developed PerturBench, a modular benchmark designed to assess predictive models of drug and gene perturbations in pharmacologically relevant contexts \cite{wu2024perturbench}.  PerturBench facilitates safer compound prioritization and enhances the robustness of dose–response outcome predictions through the use of calibration-aware metrics.Jain and Nicholls formulated guiding principles for docking benchmarks, which encompass pose accuracy, enrichment, and affinity scoring\cite{jain2008recommendations}.  These practices enhance the reliability of virtual screening funnels by reducing dataset biases and facilitating equitable comparisons of objectives. Ulman et al. established the Cell Tracking Challenge, which standardized datasets and evaluation metrics for the analysis of cell tracking and morphology \cite{mavska2014benchmark,mavska2023cell}.  These resources improve pharmacodynamic assessments and facilitate mechanism-of-action investigations by allowing for consistent image-derived phenotyping across different laboratories. Wang et al. proposed the integration of spatiotemporal cell-dynamics modeling with live-cell imaging to simulate receptor cycling, cytoskeletal reorganization, and dose–time–response relationships \cite{chen2024applying}.  Incorporating mechanistic insights into Virtual Cells reduces the experimental search space, thereby enhancing the efficiency of early-stage drug development.

\subsection{Personalized Medicine}
Virtual Cells are also advancing personalized medicine by enabling patient-specific modeling grounded in multi-omics and imaging data. Argelaguet et al. came up with guidelines for putting together scRNA, scATAC, protein, and geographical data without losing any biology\cite{argelaguet2021computational}. This is needed to make virtual cells that are accurate for each patient. Their tips on how to manage batches, align manifolds, and deal with uncertainty make sure that inferred states represent differences between patients and not technological problems. In reality, this means that model initializations are tailored to each person, which helps them make better guesses about how well therapy will work. Tim Stuart and Rahul Satija explain about integrative single-cell analysis, which creates robust joint embeddings and neighborhood graphs that can be used to model cases one at a time ~\cite{stuart2019integrative}. These embeddings make it easier to figure out the states and paths of each patient’s cells, which helps with choosing the right treatment and keeping an eye on them over time in silico. Sarah A. Teichmann and Mira Efremova examine at cross-modal single-cell approaches in both matched and mismatched scenarios. They look at selection criteria and assessment measures that stop too much alignment~\cite{todorov2019computational}. These tools produce cell-state correspondences that have been checked against each other and take uncertainty into consideration. This is especially crucial when constructing models for individual patients from data that is missing or mixed.
Todorov et al. talk about scalable, reproducible pipelines for single-cell data that can be processed quickly. These pipelines include benchmarked steps that protect rare populations that are often important for patient stratification ~\cite{todorov2019computational}. Their advice minimizes the amount of analytic drift between clinics by making preprocessing and trajectory inference more stable. This makes sure that virtual cells that are started from patient samples are the same and trustworthy. Daniel Resasco and Fei Gao et al. describe Virtual Cell’s solver stack, which includes reaction-diffusion, electrophysiology, stochastic, and rule-based layers. Clinicians and translational teams can use patient imaging and omics to set the parameters for these layers ~\cite{resasco2012virtual}. This lets you test scenarios based on your own data, like how cardiomyocytes handle ions differently or how tumors signal receptors. This helps in making therapeutic plans and predicting risks. Ion Moraru and James Schaff et al. talk about Virtual Cell’s collaborative, database-backed environment that keeps track of where data came from and makes it easier to do personalization studies across multiple centers  ~\cite{moraru2008virtual}. Across institutions, teams can share geometries, parameter sets, and validation runs. This speeds up the process of reaching agreement on patient-specific models that meet clinical QA standards. Leslie Loew and James Schaff stress the importance of making modeling from microscopy-derived geometries easy to use, which makes it easier for clinical groups to create personalized, spatially explicit models ~\cite{loew2001virtual}. When this is linked to local imaging workflows, it speedsup the process of getting patient data into in silico treatment simulations.\\ 

\subsection{Hypothesis-Driven Research}
Virtual Cells provide a computational environment for hypothesis generation, allowing researchers to explore biological mechanisms before costly experiments.Boris Slepchenkomes et al.support an image→geometry→simulation pipeline that makes predictions about transport, Ca²+  dynamics, and membrane fluxes that can be tested ~\cite{moraru2008virtual}. This completes the connection between mechanistic theory and experiment, allowing labs to test hypotheses on a computer
before doing complicated assays. R. Prill et al. (DREAM challenges) make blinded, leaderboard-based testing for systems biology models a standard
practice. This turns hypothesis generation into a community activity with strict scoring ~\cite{prill2010towards}. This culture supports virtual-cell challenges, in which mechanistic claims must hold up when tested against datasets that have not been used before. Andrew Joseph and Maya Topf et al. show how to use integrative structural modeling to combine cryo-EM, XL-MS, and proteomics to make ensembles with clear uncertainty, which adds physically plausible constraints to hypothesis tests ~\cite{malhotra2021assessment}. Virtual cells that are seeded with these shapes can tell the difference between competing hypotheses about how things should be put together or where they should be located.Tyson et al.  created dynamic models to analyze estrogen receptor (ER) signaling within breast cancer cells, emphasizing the interconnected relationships among cell cycle progression, apoptosis, autophagy, and the unfolded protein response.  Through the integration of these pathways, the study demonstrated the potential of computational modeling to forecast cell-fate decisions across various therapeutic scenarios.  The research emphasized the mechanisms that regulate endocrine responsiveness and resistance, demonstrating the significance of systems-level models in oncology.    This method utilizes the concept of Virtual Cells, employing predictive simulations to generate hypotheses and treatment plans.    These applications demonstrate the significant potential of Virtual Cells in enhancing personalized medicine and targeted cancer therapy\cite{tyson2011dynamic}. Burkhardt et al. provide a comprehensive review of the ways in which phenotypic plasticity influences cancer progression and therapy resistance, emphasizing the significance of manifold learning as an effective analytical framework \cite{burkhardt2022mapping}.  The authors present a framework for understanding cancer cells as existing within a continuous state landscape, demonstrating how non-genetic transitions—like epithelial-to-mesenchymal shifts, metabolic rewiring, and immune evasion—contribute to the diversity observed within tumors.  It is highlighted that the integration of single-cell omics technologies with manifold learning has the potential to effectively capture the overall structure and specific pathways of cell state transitions.  This method helps identify stable attractor states and canalizing traits that define plasticity pathways.   The research underscores that mapping these landscapes can facilitate the formulation of therapies designed to avert the transition into resistant states, thereby offering novel avenues to diminish metastasis and improve treatment outcomes.Rui Alves et al. show how tool chaining and ontology alignment turn messy database evidence into network drafts that can be used to improve specific hypotheses ~\cite{karathia2011saccharomyces}. The workflow shows which experiments to run to figure out the mechanism by showing where the conflict points and origins are. Trey Ideker et al. say that scaffold-centric integration—pathways, complexes, and compartments—should be the common language for comparing mechanistic hypotheses across datasets ~\cite{ideker2003building}. This lets labs make testable modules and keep track of their progress as more evidence comes in. viceconti et al. write down multiscale verification/validation so that hypotheses that  go from molecule to cell to tissue can be tested with the right connections and error budgets ~\cite{viceconti2008virtual}. Their engineering discipline makes people less sure of themselves when results depend on cross-scale interactions that are often missed in informal analyses.

\section{Challenges and Open Problems}

The examination of Virtual Cells is characterized by both innovation and ongoing challenges that influence their development.  Researchers working to enhance these models for biological accuracy encounter challenges associated with data curation, computational scalability, reproducibility, and platform interoperability. Resolving these issues is crucial for enhancing Virtual Cells as effective instruments for biomedical research and application.Fahsbender and Andersson established the Virtual Cell Challenge as an international standard for evaluating predictive models of cellular behavior \cite{fahsbender2025benchmarking}.  The challenge highlighted the significance of fair comparison through the use of hidden test sets, reproducible leaderboards, and evaluation protocols centered on generalization.  The text emphasizes the necessity of uncertainty reporting, provenance tracking, and management of domain shifts, illustrating the variability present among laboratories, assays, and species.  Despite advancements, the generation of gold-standard datasets continues to be slow and resource-intensive, especially when integrating clinical data constrained by privacy regulations.  Computational inequities persist, as confined laboratories frequently lack access to high-performance resources, which raises concerns regarding fairness.  Schnabel and Davatzikos highlighted the significance of dependable imaging benchmarks, stressing the need for clear tasks, statistical rigor, and the maintenance of sustainable datasets \cite{resasco2012virtual}.  These studies highlight that reproducibility and fairness are critical yet unresolved challenges. Whole-cell models necessitate high-quality, integrative data.  Macklin et al. observed that existing datasets frequently exhibit issues such as being siloed, incomplete, or inadequately annotated, which complicates calibration and validation processes \cite{karr2012whole}.  Models are capable of interpolating between established conditions; however, they often struggle to generalize beyond these parameters.  The proposed solution of integrating experimental design, parameter inference, and validation into a continuous pipeline is still aspirational.  Imaging-derived phenotypes also show problems with annotation.   Maška and Ulman addressed this problem with the Cell tracing Challenge, which created defined standards and criteria for lineage tracing \cite{mavska2023cell}. Expert annotations are quite expensive, and the description of unusual cellular activities is currently not good enough.   Uncertainty-aware tracking and automated labeling show promise for progress, although they are yet not fully developed0\\.

Integrating imaging data with mechanistic models continues to pose a significant challenge.  Moraru et al. illustrated the conversion of microscopy-derived geometries into computational domains suitable for ordinary differential equations (ODE), partial differential equations (PDE), and stochastic solvers, thereby transforming static images into predictive experiments \cite{moraru2008virtual}.  This method identified significant challenges, such as the reconstruction of precise 3D geometries from noisy images, artifact-free mesh generation, and the integration of subcellular electrical activity.  Slepchenko et al. emphasized the significance of spatial organization, demonstrating that reaction–diffusion and compartmental models can influence biological outcomes, including calcium propagation and signaling dynamics \cite{slepchenko2003quantitative}  However, these advancements are counterbalanced by practical limitations: the quality of the mesh directly influences accuracy, solvers encounter difficulties with stiffness, and numerous kinetic parameters remain unidentifiable.  The assessment of the reliability of spatially explicit models is challenging in the absence of high-resolution validation data. Loew and Schaff developed the VCell platform, offering a client–server environment for model construction and simulation without the necessity of coding \cite{loew2001virtual}.   Simulation protocols often showed a tendency towards platform specificity, with stochastic replicates and parameter sweeps revealing restricted portability, while meshing artifacts were still apparent in intricate geometries.   The later developments by Moraru and colleagues enabled the joint use of shared geometries, simulation histories, and databases with version control \cite{moraru2008virtual}.   This encouraged cooperation among various institutions; however, discrepancies in metadata and significant computational requirements persisted as obstacles.  Proposals for the packaging of simulations as portable artifacts have been presented \cite{lu2019integrated}; however, a deficiency in widely accepted standards persists. Challenges including data privacy issues, variations in computational resources, and the absence of interoperable standards hinder progress across various domains.The FAIR principles—Findable, Accessible, Interoperable, Reusable—and community schemas for compartments, membranes, and simulation protocols are increasingly acknowledged as essential. However, their implementation remains inconsistent.  Considerations of ethics involve the imperative to protect patient privacy within tailored models and the critical need to avoid the marginalization of under-resourced groups to promote fair adoption. The challenges in this field indicates that Virtual Cells have evolved from theoretical models to data-driven, collaborative ecosystems, yet they continue to encounter issues with reproducibility, scalability, and standardization.  Every challenge in data annotation, solver performance, or benchmark design represents a possible path for advancement.

\begin{itemize}
\item Data Generation
\item Matrix and Evaluation
\item Biological interpretability
\item Standardization of modeling frameworks
\item Computational scalability
\end{itemize}

\section{Future Outlook}

The advancement of Virtual Cell (VCell) studies will hinge on the community's ability to tackle current obstacles while leveraging previous achievements.  VCells are evolving from specialized computational tools into essential platforms for biomedical discovery, clinical translation, and personalized medicine.  The advancements in data integration, benchmarking, and computational modeling are encouraging; however, significant gaps still exist.  Future development will necessitate standardized datasets, transparent and interpretable models, user-friendly and interoperable platforms, scalable computational methods, and robust global collaboration backed by open science practices.
One of the main objective is to produce large, high-quality, and standardized datasets.  The Virtual Cell Challenge emphasizes the necessity for simulations to incorporate data from diverse domains such as genomics, transcriptomics, proteomics, metabolomics, live-cell imaging, and electrophysiology, all of which must be gathered in a controlled environment~\cite{schaff2012virtual}. In the absence of integrated resources, models tend to overfit and struggle to generalize effectively. Initiatives like PerturBench~\cite{wu2024perturbench} and the DREAM challenges emphasize the importance of organized data sharing and reproducible benchmarking.  Significant gaps exist in time-series multi-omics that include perturbations, clinically matched patient datasets, and high-resolution 3D reconstructions of diverse cell types.   Federated learning and privacy-preserving computation can facilitate the integration of clinical data while maintaining confidentiality. Understanding model interpretability is a significant focus area.  The VCell platform has consistently focused on providing clear biochemical representation and spatial compartmentalization~\cite{loew2001virtual,schaff2016rule}. As models grow more intricate, apprehensions about possible "black-box" behavior are escalating.   Waltemath et al. emphasize the importance of modular design, sensitivity analysis, and semantic annotation in achieving clarity.~\cite{waltemath2011reproducible}. New methods from explainable AI, such causal graph inference and feature attribution, might make it easier to combine mechanical and statistical modeling.   Future platforms may provide interactive dashboards that link predictions with molecular interactions, parameters, and references, thereby improving trust between computational and experimental scientists. Usability and interoperability are essential considerations.  Despite advancements in solver engines and architectures~\cite{Leo2024ICAICA,Leo2024MedDoc}, numerous systems continue to pose challenges for non-experts.  Macklin~\cite{schaff2012virtual} and Resasco~\cite{resasco2012virtual} emphasize the significance of standardized APIs, exchange formats, and modular plugins for the integration of imaging pipelines and laboratory systems.  Cloud-native collaborative platforms are expected to prevail in the future, facilitating real-time large-scale simulations without the need for local high-performance computing resources.  AI-assisted model building has the potential to lower barriers by automating the processes of geometry reconstruction, parameter estimation, and pathway design.\\

Computational scalability represents a significant challenge.  High-resolution 3D and multi-scale models continue to require significant resources.  Moraru et al. and Slepchenko et al. emphasize that adaptive solvers, GPU/TPU acceleration, and surrogate modeling are noteworthy solutions. ~\cite{moraru2008virtual,slepchenko2003quantitative}.   Adaptive resolution modeling allocates increased precision to critical regions, such as signaling domains, while optimizing less significant areas, potentially facilitating a balance between performance and fidelity.The ongoing achievements of Virtual Cells depend on global partnerships and the principles of open science.     Collaborative benchmarks, shared repositories, and partnerships between institutions have significantly accelerated progress~\cite{wu2025individual}.     Enhancing these initiatives will facilitate access, minimize redundancy, and promote fair participation across various resource settings. Prioritizing ethical considerations, including fairness, privacy, and accountability, is essential when VCells are utilized in clinical settings.Future prospects indicate several transformative directions.  These encompass "living models" that are perpetually updated with new data, cross-scale digital twins that connect cells with organs and tissues, and their incorporation into clinical decision support systems.  Enhancing explainability, usability, benchmarking, and scalability may enable Virtual Cells to transition from research prototypes to reliable instruments for biomedical discovery and personalized therapy.

\begin{itemize}
\item Expanding high-quality, standardized datasets
\item Enhancing interpretability tools
\item Developing integrative, user-friendly modeling platforms
\end{itemize}

\section{Conclusions}
Virtual cells have progressed from ambitious demonstrations to credible, reusable infrastructure for mechanistic discovery, prediction, and design. Early platforms such as VCell established the core blueprint: unite reaction–kinetics, transport, and electrophysiology in spatially realistic geometries, expose these capabilities through user-facing workflows, and preserve provenance so that results are auditable and sharable. That blueprint has expanded in step with the measurement revolution. Single-cell and spatial assays, high-content imaging, and integrative structural biology now provide the priors, constraints, and boundary conditions required to initialize and validate models that explicitly account for heterogeneity, compartmentalization, and geometry-dependent dynamics. In parallel, multiscale formalisms and co-simulation strategies connect cellular mechanisms to tissue- and organ-level behavior, clarifying how local biophysics scales to physiological function~\cite{terrell2022management}.

A central theme in this review is hybridization: mechanistic formalisms (ODE/PDE/stochastic, rule-based kinetics, and continuum or agent-based mechanics) increasingly cooperate with machine learning components for representation learning, parameter inference, emulator construction, and uncertainty quantification. This synergy is pragmatic rather than ideological. Mechanistic models encode causal structure, constraints, and interpretability; data-driven modules accelerate expensive solvers, extract informative latent variables from noisy modalities, and supply priors when parameters are weakly identifiable. Together they enable \emph{living} models that co-evolve with data streams and knowledge graphs, improving predictions and guiding experiment through design–build–test–learn loops.

Despite this momentum, several challenges must be addressed to mainstream virtual cells. First, \textbf{identifiability and uncertainty}: spatial models with many species and processes are underdetermined by typical datasets. Robust priors, targeted experiments, and posterior diagnostics should become routine, with uncertainty reported as a first-class output rather than a footnote. Second, \textbf{computational scale}: fully 3D geometries, wide time-scale separations, and hybrid stochastic–deterministic dynamics strain even modern hardware. Continued progress in adaptive discretization, GPU acceleration, and surrogate modeling will expand the feasible frontier. Third, \textbf{standards and interoperability}: exchangeable model and data formats, curated ontologies, and reference benchmarks are essential for reproducibility, fair comparison, and education. Lessons from earlier software ecosystems (e.g., E-Cell and scaffold-based integration) underscore the value of modularity, clear interfaces, and versioned artifacts~\cite{nguyen2024absence}.

We also emphasize \textbf{human factors}. The most impactful platforms do not sacrifice rigor for usability; they pair mathematically sound solvers with literate workflows, visual inspection tools, and guardrails for good practice. As community adoption grows, documentation, exemplars, and teaching materials will matter as much as algorithmic novelty~\cite{zhu2024sni}. In the same spirit, reference use-cases---calcium waves in realistic morphologies, nucleocytoplasmic transport with explicit barriers, or ion homeostasis on complex membranes---should serve as shared \emph{competence tests} that validate end-to-end pipelines from imaging to simulation to explanation ~\cite{liu2023regformer}.

Looking forward, we anticipate three convergences. (\emph{i}) \emph{Data-model co-design}: perturbational single-cell/spatial assays will be planned explicitly to resolve model uncertainties and validate counterfactual predictions, tightening the experimental--computational loop. (\emph{ii}) \emph{Structure-to-function integration}: mesoscale structural ensembles will increasingly furnish geometry and restraints for reaction–diffusion and mechanochemical models, shortening the path from map to mechanism. (\emph{iii}) \emph{Cross-scale composition}: virtual cells will plug into multiscale physiological contexts via standardized interfaces, enabling principled upscaling of cellular dynamics to tissue and organ function. As precedents such as the first whole-cell models show, end-to-end coordination across submodels is feasible and scientifically generative, even if today’s eukaryotic scale remains aspirational~\cite{Zhang2025TimeLLaMA}.

In sum, virtual cells are poised to become a common language between quantitative theory and experiment. Their value rests not only on accurate predictions, but on their ability to reveal \emph{why} systems behave as observed, to suggest informative experiments, and to provide transparent, reusable artifacts that other groups can scrutinize and extend. Achieving this vision will require sustained attention to standards, benchmarks, and user-centered design as much as to new algorithms. With these commitments, virtual cells can transition from specialist tools to broadly useful scientific infrastructure---connecting molecular events to cellular phenotypes and, ultimately, to physiology and therapy.

\bibliographystyle{plain}
\bibliography{reference}

\end{document}